\documentclass[aps,prd,a4paper,twocolumn,amsmath,showpacs,superscriptaddress,nofootinbib,preprintnumbers]{revtex4-1}
\pdfoutput=1

\usepackage{amssymb,amsmath,latexsym,mathrsfs}
\usepackage[sort&compress]{natbib}
\usepackage{graphicx,subfigure}
\usepackage{epsfig}
\usepackage{varioref,xr-hyper}
\usepackage{color}
\usepackage{multirow}
\usepackage{array}
\usepackage{hyperref}
\usepackage{wasysym}
\usepackage{color}
\usepackage{float}
\usepackage{xcolor}
\usepackage[utf8]{inputenc}
\usepackage[T1]{fontenc}
\usepackage[sort&compress]{natbib}

\begin{document}

\title{Listening to the sound of dark sector interactions with gravitational wave standard sirens}

\author{Weiqiang Yang}
\email{d11102004@163.com}
\affiliation{Department of Physics, Liaoning Normal University, Dalian, 116029, P. R. China}

\author{Sunny Vagnozzi}
\email{sunny.vagnozzi@fysik.su.se}
\affiliation{The Oskar Klein Centre for Cosmoparticle Physics, Stockholm University, Roslagstullsbacken 21A, SE-106 91 Stockholm, Sweden}
\affiliation{The Nordic Institute for Theoretical Physics (NORDITA), Roslagstullsbacken 23, SE-106 91 Stockholm, Sweden}
\affiliation{Kavli Institute for Cosmology (KICC) and Institute of Astronomy, University of Cambridge, Madingley Road, Cambridge CB3 0HA, United Kingdom}

\author{Eleonora Di Valentino}
\email{eleonora.divalentino@manchester.ac.uk}
\affiliation{Jodrell Bank Center for Astrophysics, School of Physics and Astronomy, University of Manchester, Oxford Road, Manchester M13 9PL, United Kingdom}

\author{Rafael C. Nunes}
\email{rafadcnunes@gmail.com}
\affiliation{Divis\~{a}o de Astrof\'{i}sica, Instituto Nacional de Pesquisas Espaciais, Avenida dos Astronautas 1758, S\~{a}o Jos\'{e} dos Campos, 12227-010, SP, Brazil}

\author{Supriya Pan}
\email{supriya.maths@presiuniv.ac.in}
\affiliation{Department of Mathematics, Presidency University, 86/1 College Street, Kolkata 700073, India}

\author{David F. Mota}
\email{d.f.mota@astro.uio.no}
\affiliation{Institute of Theoretical Astrophysics, University of Oslo, P.O. Box 1029 Blindern, N-0315 Oslo, Norway}

\begin{abstract}
We consider two stable Interacting Dark Matter -- Dark Energy models and confront them against current Cosmic Microwave Background data from the \textit{Planck} satellite. We then generate luminosity distance measurements from ${\cal O}(10^3)$ mock Gravitational Wave events matching the expected sensitivity of the proposed Einstein Telescope. We use these to forecast how the addition of Gravitational Wave standard sirens data can improve current limits on the Dark Matter -- Dark Energy coupling strength ($\xi$). We find that the addition of Gravitational Waves data can reduce the current uncertainty by a factor of $5$. Moreover, if the underlying cosmological model truly features Dark Matter -- Dark Energy interactions with a value of $\xi$ within the currently allowed $1\sigma$ upper limit, the addition of Gravitational Wave data would help disentangle such an interaction from the standard case of no interaction at a significance of more than $3\sigma$.
\end{abstract}

\pacs{98.80.-k, 98.80.Cq, 95.35.+d, 95.36.+x, 98.80.Es.}
\maketitle
 
\section{Introduction}
\label{sec:intro}

Dark matter (DM) and dark energy (DE), despite their unknown nature, are two key ingredients of the standard cosmological model. Within the concordance $\Lambda$CDM cosmological model, widely supported by a number of independent observations (for instance~\cite{Riess:1998cb,Perlmutter:1998np,Alam:2016hwk,Abbott:2017wau,Aghanim:2018eyx}), the role of DM is played by a cold pressureless fluid, whereas the role of DE is played by a cosmological constant. In the standard picture, DM and DE evolve separately, each obeying a separate continuity equation, and do not interact if not gravitationally. However, from a microphysical perspective, most field theoretical descriptions of DM and DE lead to interactions between the two. In fact, even if absent at tree level, a DM-DE interaction will inevitably be generated at loop level if not explicitly forbidden by a fundamental symmetry~\cite{Wetterich:1987fm}.~\footnote{See e.g.~\cite{Amendola:2007yx,Micheletti:2009pk,Pavan:2011xn,Bolotin:2013jpa,Costa:2014pba} for examples of explicit particle realizations of DM-DE interactions. Moreover, a number of modified gravity models can be recasted as interacting DM-DE models when expressed in the Einstein frame. See e.g.~\cite{Poplawski:2006kv,He:2011qn,He:2012rf} for the case of $f(R)$ gravity~\cite{Nojiri:2006ri,DeFelice:2010aj,Nojiri:2010wj}, e.g.~\cite{Dutta:2017fjw} for the case of mimetic gravity and variants thereof~\cite{Chamseddine:2013kea,Myrzakulov:2015qaa,Cognola:2016gjy,Sebastiani:2016ras,Vagnozzi:2017ilo,deHaro:2018sqw}, and e.g.~\cite{Koivisto:2012za,Chiba:2013mha,Koivisto:2015qua,Kofinas:2016fcp,Dutta:2017wfd,Zonunmawia:2017ofc,Bahamonde:2017ize} for discussions in the context of other theories.} There exist a plethora of models featuring DM-DE interactions, usually referred to as interacting dark energy (IDE) models, most of which phenomenological in nature. For an incomplete selection of works examining IDE models from the model-building, theoretical, and observational perspectives, see for instance~\cite{Amendola:1999er,Billyard:2000bh,Barrow:2002zh,Farrar:2003uw,Skordis:2005xk,Barrow:2006hia,Amendola:2006dg,He:2008tn,Nunes:2014mt,Quartin:2008px,Valiviita:2008iv,Chen:2008ft,CalderaCabral:2008bx,delCampo:2008jx,Jackson:2009mz,CalderaCabral:2009ja,Majerotto:2009np,Jamil:2009eb,Abdalla:2009mt,Martinelli:2010rt,Honorez:2010rr,DeBernardis:2011iw,Baldi:2011qi,Clemson:2011an,Pan:2012ki,Pan:2013rha,Otalora:2013tba,Costa:2013sva,Li:2013bya,Yang:2014vza,Yang:2014gza,Faraoni:2014vra,Pan:2014afa,Abdalla:2014cla,Gleyzes:2015pma,Tamanini:2015iia,Marra:2015iwa,Li:2015vla,Murgia:2016ccp,Nunes:2016dlj,Costa:2016tpb,Marcondes:2016reb,Tutusaus:2016kyl,Kumar:2016zpg,Xia:2016vnp,Yang:2016evp,Pan:2016ngu,Marttens:2016cba,Mukherjee:2016shl,Sharov:2017iue,Kumar:2017dnp,DiValentino:2017iww,An:2017kqu,Santos:2017bqm,Mifsud:2017fsy,Kumar:2017bpv,Guo:2017deu,Pan:2017ent,An:2017crg,Feng:2017usu,Guo:2018gyo,Costa:2018aoy,Cardenas:2018qcg,Yang:2018pej,Wang:2018azy,vonMarttens:2018iav,Yang:2018xlt,Elizalde:2018ahd,Yang:2018uae,Zhang:2018glx,LeDelliou:2018vua,Li:2018ydj,vonMarttens:2018bvz,Cardenas:2018nem,Yang:2018qec,Bonici:2018qli,Costa:2019uvk,Mifsud:2019hjw,Martinelli:2019dau,Paliathanasis:2019hbi,Kumar:2019wfs,Feng:2019mym,Pan:2019jqh,Li:2019loh,Dai:2019vif}. For a recent comprehensive review on IDE models, see~\cite{Wang:2016lxa}.

In 2015, the detection of the gravitational wave (GW) event GW150914~\cite{Abbott:2016blz} by the LIGO collaboration officially inaugurated the era of GW astronomy, which has opened an unprecedented window onto tests of fundamental physics~\cite{Maggiore:1999vm,Schutz:1999xj,Berti:2015itd}. In particular, the detection of the GW event GW170817~\cite{TheLIGOScientific:2017qsa} and its electromagnetic (EM) counterpart GRB170817A~\cite{Monitor:2017mdv} has marked the dawn of the multi-messenger astronomy era, and has already been successfully utilized to place tight constraints on various aspects of fundamental physics, see e.g.~\cite{Creminelli:2017sry,Sakstein:2017xjx,Ezquiaga:2017ekz,Baker:2017hug,Boran:2017rdn,Nojiri:2017hai,Arai:2017hxj,Amendola:2017orw,Visinelli:2017bny,Crisostomi:2017lbg,Langlois:2017dyl,Gumrukcuoglu:2017ijh,Kreisch:2017uet,Addazi:2017nmg,Dima:2017pwp,Cai:2018rzd,Pardo:2018ipy,Casalino:2018tcd,Wan:2018udw,Jana:2018djs,Ganz:2018vzg,Casalino:2018wnc,Cai:2019jah} (see also the reviews~\cite{Barack:2018yly,Ezquiaga:2018btd}). 

One particularly intriguing use of GWs is the possibility of exploiting them as \textit{standard sirens} (SS)~\cite{Holz:2005df,Palmese:2019ehe} (see e.g. Chapter 1, Sec.~13 of~\cite{Barack:2018yly} for a review), a possibility first proposed in~\cite{Schutz:1986gp} to measure the Hubble constant $H_0$. In fact, an accurately measured GW signal allows one to reconstruct the luminosity distance to the source $d_L$, and hence can be used as a distance indicator. This possibility was in fact first exploited in~\cite{Abbott:2017xzu} to provide the first ever SS measurement of $H_0$,  and for the first time this new measurement was combined with the \textit{Planck} Cosmic Microwave Background (CMB) data in \cite{DiValentino:2017clw}. The SS measurement of $d_L$ is best performed if the observation of an EM counterpart allows one to determine the redshift $z$ of the source~\cite{Schutz:1986gp}, but is in principle possible through a more complicated statistical approach even if direct redshift information is lacking~\cite{Schutz:1986gp,MacLeod:2007jd,DelPozzo:2011yh} (although such procedure is not free from complications, see for instance~\cite{Messenger:2012jy}).

The possibility of using SS to constrain the late-time dynamics of the Universe, including the Hubble constant $H_0$, the matter density parameter $\Omega_{m0}$, and the dark energy equation of state $w_x$, has been contemplated in a number of works: for an incomplete list, see e.g.~\cite{Dalal:2006qt,Stavridis:2009ys,Petiteau:2011we,Nishizawa:2016ood,Seto:2017swx,Chen:2017rfc,Feeney:2018mkj,Wang:2018lun,DiValentino:2018jbh,Du:2018tia,Nunes:2019bjq}. Moreover, the first work combining the GWs probe and particle collider constraints was put forward in \cite{Huang:2016odd}.  In particular, the work of~\cite{Cai:2017yww} first explored the possibility of reconstructing interactions between DM and DE using future data from LISA. 

In this work, it is our goal to revisit the possibility of using GW SS measurements to study interactions between DM and DE. Basing ourselves on the formalism of~\cite{Du:2018tia} (see also~\cite{Yang:2019bpr,Zhang:2019ple,Bachega:2019fki} for later work), we explore how the use of future GW data can improve constraints on DM-DE interactions, within the context of two stable interacting DE models proposed by the present authors in~\cite{Yang:2018euj}.

This paper is then organized as follows. In Sec.~\ref{sec:backgroundperturbations} we provide a very brief overview of IDE models, focusing in particular on the two IDE models presented in~\cite{Yang:2018euj} and which we will consider in this work. In Sec.~\ref{sec:data} we describe the data we will make use of in this work: we consider both current data in the form of CMB anisotropy and Baryon Acoustic Oscillation (BAO) distance measurements, as well as mock luminosity distance measurements from future GW standard sirens data, and briefly describe our methodology for generating the mock data. In Sec.~\ref{sec:results} we present our results forecasting the ability of future GW data to improve current constraints on the IDE models under consideration. We conclude with closing remarks in Sec.~\ref{sec:conclusions}.

\section{Stable interacting dark energy revisited}
\label{sec:backgroundperturbations}

In the following, we provide the basic equations describing the evolution of the Universe in presence of DM-DE interactions. As usual, we work within the framework of a homogeneous and isotropic Universe described by a spatially flat Friedmann-Lema\^{i}tre-Robertson-Walker (FLRW) line element. We take the gravitational sector of the Universe to be described by Einstein's theory of General Relativity, whereas we consider a matter sector minimally coupled to gravity. The energy budget of the Universe is provided by five species: baryons, photons, neutrinos, dark matter (DM) and dark energy (DE).~\footnote{We fix the total neutrino mass to $M_{\nu}=0.06\,{\rm eV}$, as done in the \textit{Planck} baseline analyses. Given the currently very tight upper limits on $M_{\nu}$, of order $0.1\,{\rm eV}$~\cite{Palanque-Delabrouille:2015pga,Giusarma:2016phn,Vagnozzi:2017ovm,Giusarma:2018jei,Aghanim:2018eyx}, we do not expect the introduction of massive neutrinos to the picture will change our conclusions substantially.} In particular, we allow DM and DE to interact, with the specific form of the interaction to be described shortly.

In the absence of interactions, the energy-momentum tensors of the DM and DE components are separately conserved, \textit{i.e.} $\nabla_{\mu}T_i^{\mu \nu}=0$, where $i$ stands for DM or DE. In the presence of non-gravitational DM-DE interactions, a convenient phenomenological parametrization of the effect of these interactions is obtained by modifying the conservation equations for the stress-energy tensors as follows~\cite{Valiviita:2008iv,Majerotto:2009np,Clemson:2011an,Gavela:2009cy,Gavela:2010tm}:
\begin{eqnarray}
\nabla_{\mu}T_{i}^{\mu \nu }=Q_{i}^{\nu}\,, \quad \sum\limits_{\mathrm{i}}{%
Q_{i}^{\mu }}=0,
\end{eqnarray}
where again $i$ stands for either DM ($i=c$) or DE ($i=x$). The four-vector $Q_{i}^{\mu}$ specifies the coupling between the dark sectors, and characterizes the type of interaction. We take $Q_{i}^{\mu}$ to assume the following form:
\begin{eqnarray}
Q_{i}^{\mu}=(Q_{i}+\delta Q_{i})u^{\mu}+a^{-1}(0,\partial^{\mu}f_{i}), 
\end{eqnarray}
where $u^{\mu}$ is the velocity four-vector, $Q_i$ is the background energy transfer, and $f_i$ is the momentum transfer potential. From now on, we take $Q_i \equiv Q$. In addition, we consider the simplest possibility wherein the momentum transfer potential is zero in the rest frame of DM, see e.g.~\cite{Valiviita:2008iv,Majerotto:2009np,Clemson:2011an,Gavela:2009cy,Gavela:2010tm} for more details.

All that is left, therefore, is to specify the functional form of $Q$. In what follows, we will consider a particular class of IDE models, whose appeal is their being free from early-time linear perturbation instabilities. These models were proposed by us in~\cite{Yang:2018euj}, and we refer the reader there for more details. We consider the same models, keeping the same labels \texttt{IDErc1} and \texttt{IDErc2} respectively, whose coupling functions $Q$ take the form~\cite{Yang:2018euj}:
\begin{eqnarray}
\label{model1}
Q&=&3 (1+w_x) H \xi \rho_c \quad (\texttt{IDErc1})\,, \\
\label{model2}
Q&=&3(1+w_x)H\xi(\rho_c+\rho_x) \quad (\texttt{IDErc2})\,.
\end{eqnarray}

In Eqs.~(\ref{model1},\ref{model2}), $w_x$ denotes the DE equation of state (EoS), $H$ is the Hubble expansion rate, whereas $\rho_c$ and $\rho_x$ are the DM and DE energy densities respectively. Finally, the parameter $\xi$ controls the strength of the DM-DE coupling, with $\xi=0$ representing the standard case of no interaction. The introduction of the factor $(1+w_x)$ in Eqs.~(\ref{model1},\ref{model2}), absent in earlier IDE works, allows for stable early-time perturbations \textit{independently of the value of $w_x$} (whereas earlier models usually featured stable perturbations only for either quintessence-like or phantom DE EoS, limiting the possibilities in terms of DE fluid allowed to interact), see~\cite{Yang:2017zjs,Yang:2017ccc,Yang:2018euj} for further discussions. In particular, as shown in~\cite{Yang:2018euj}, both the \texttt{IDErc1} and \texttt{IDErc2} models are stable for $\xi>0$. Our aim is to explore how the addition of future luminosity distance measurements from GW standard sirens can improve constraints from current cosmological data on the DM-DE coupling strength $\xi$.

\section{Observational data and methodology}
\label{sec:data}

In the following, we describe in more detail the observational datasets (both current and future) we include in this work. In terms of current observational datasets, we make use of measurements of Cosmic Microwave Background (CMB) temperature and polarization anisotropies and their cross-correlations from the \textit{Planck} 2015 data release~\cite{Ade:2015xua}. We analyse these measurements making use of the publicly available \textit{Planck} likelihood~\cite{Aghanim:2015xee}. We refer to this dataset as ``CMB'' (note that this dataset is usually referred to as \textit{PlanckTTTEEE}+\textit{lowTEB} in the literature).

To forecast the constraining ability of future GW standard sirens (SS) luminosity distance measurements, we generate mock data matching the expected sensitivity of the Einstein Telescope. The Einstein Telescope is a proposed third-generation ground-based GW detector, whose main objectives will be to test General Relativity in the strong field regime and advance precision GW astronomy~\cite{Sathyaprakash:2012jk}. Although full instrumental details are still under study, the Einstein Telescope will likely be located underground and will feature three ${\cal O}(10\,{\rm km})$ long arms arranged following an equilateral triangle. Each arm will be composed of two interferometers, optimized for operating at frequencies of ${\cal O}(1-100)\,{\rm Hz}$ and ${\cal O}(0.1-100)\,{\rm kHz}$ respectively~\cite{Sathyaprakash:2012jk}. After 10 years of operation, the Einstein Telescope is expected to detect ${\cal O}(10^3)$ GW SS events.

Our goal is then to generate a luminosity distance catalogue matching the expected sensitivity of the Einstein Telescope after 10 years of operation. We generate $1000$ triples $(z_i,d_{L}(z_i),\sigma_i)$, with $z_i$ the redshift of the GW source, $d_L$ the measured luminosity distance, and $\sigma_i$ the uncertainty on the latter. There are three aspects to take into consideration when generating this mock data: the fiducial cosmological model enters both in $z_i$ (or more precisely into the redshift distribution of expected sources) and $d_L$, the expected type of GW sources enter in $z_i$, and finally the instrumental specifications enter in $\sigma_i$.

We now very briefly summarize the procedure adopted for generating the mock GW data and further details on the generation of the mock GW standard sirens dataset are presented in Appendix~A. In addition, we encourage the reader to consult~\cite{Du:2018tia} for further technical details on the procedure, which is the same as that adopted here. The first step is to specify the expected GW sources. We consider a combination of black hole-neutron star and binary neutron star mergers. The ratio of number of events for the former versus the latter is taken to be $0.03$, with mass distributions specified in~\cite{Du:2018tia}. Following~\cite{Schneider:2000sg,Cutler:2009qv,Cai:2016sby,Cai:2017aea,Du:2018tia}, we then model the merger rate of the sources $R(z)$, and from their merger rate we are able to determine their redshift distribution $P(z)$ (see~\cite{Du:2018tia} for detailed formulas). Once we have $P(z)$, we sample $1000$ values of redshifts from this distribution: these will be the redshifts of our mock GW events $z_i$.

Note that going from merger rate to redshift distribution requires a choice of fiducial cosmological model, as the expression for $P(z)$ contains both the comoving distance and expansion rate at redshift $z$, $\chi(z)$ and $H(z)$ respectively (see~\cite{Du:2018tia}). Since our goal is to explore how GW data can improve our constraints on IDE models, we generate two different mock GW datasets choosing as fiducial models first the \texttt{IDErc1} and then the \texttt{IDErc2} models. We adopt fiducial values for the cosmological parameters given by the best-fit values of the same parameters obtained analysing the models with the CMB dataset previously described. Using the same choice of fiducial model(s) and parameters, and the values of redshifts we sampled from $P(z)$, we then compute the luminosity distances at the respective redshifts, $d_L(z_i)$, through:
\begin{eqnarray}
d_L(z_i) = (1+z_i)\int_0^{z_i}\frac{dz'}{H(z')}\,.
\label{eq:luminosity}
\end{eqnarray}

Having obtained $z_i$ and $d_L(z_i)$, all that is left to complete our mock GW luminosity distances catalogue is $\sigma_i$. We determine these error bars following the Fisher matrix approach outlined in~\cite{Du:2018tia}. In summary, this is achieved by modelling the observed GW strain as linear combination of the two GW polarizations weighted by the two beam pattern functions. Instrumental specifications enter in determining exact functional form of the beam pattern functions. We follow~\cite{Zhao:2010sz,Cai:2016sby,Wang:2018lun} in determining the functional form of the beam pattern functions for the Einstein Telescope. We then compute the Fourier transform of the observed GW strain, $h(f)$.  In particular, the phase and amplitude of the Fourier transform (in frequency domain) are estimated for the post-Newtonian waveform TaylorF2 at 3.5 post-Newtonian order (3.5 PN in the standard notation). When generating the data, during the integration of the signal-to-noise ratio we assume a minimum frequency $f_{\min} = 1\,{\rm Hz}$ and a maximum frequency $f_{\max} = 2f_{\rm isco}$, where $f_{\rm isco}$ is the  orbital frequency of the last stable orbit associated with each simulated event. 

The crucial point to note is that amplitude of the GW signal is inversely proportional to the distance luminosity distance $d_L(z)$ (see appendix~A). Thus, assuming that the errors on $d_L(z)$ are uncorrelated with errors on the remaining waveform parameters (which is true if the distance to source does not correlate with other parameters), it is possible to show that the instrumental uncertainty on the luminosity distance is given by $\sigma^{\rm ins}_{d_L(z_i)} \propto d_L(z_i)/{\rm SNR}_{i}$, where ${\rm SNR}$ is the signal-to-noise ratio associated to the $i$th event (see Appendix~A for further details on the estimation of ${\rm SNR}$ and the instrumental uncertainty).

In addition to instrumental uncertainties, GWs are also lensed along their journey from the source to us. This results in an additional lensing uncertainty which following~\cite{Cai:2016sby} we take to be $0.05z_id_L(z_i)$. Adding in quadrature the instrumental and lensing errors we obtain the total error $\sigma_i$. Finally, we randomly displace the previously determined luminosity distances $d_L(z_i)$ by quantities $\Delta_i \sim {\cal N}(0,\sigma_i)$, \textit{i.e.} the displacements are drawn from normal distributions with mean $0$ and standard deviation $\sigma_i$. For further details, we invite the reader to consult~\cite{Du:2018tia}. The resulting mock dataset is referred to as ``GW''.

\begin{figure*}
\centering
\includegraphics[width=0.4\textwidth]{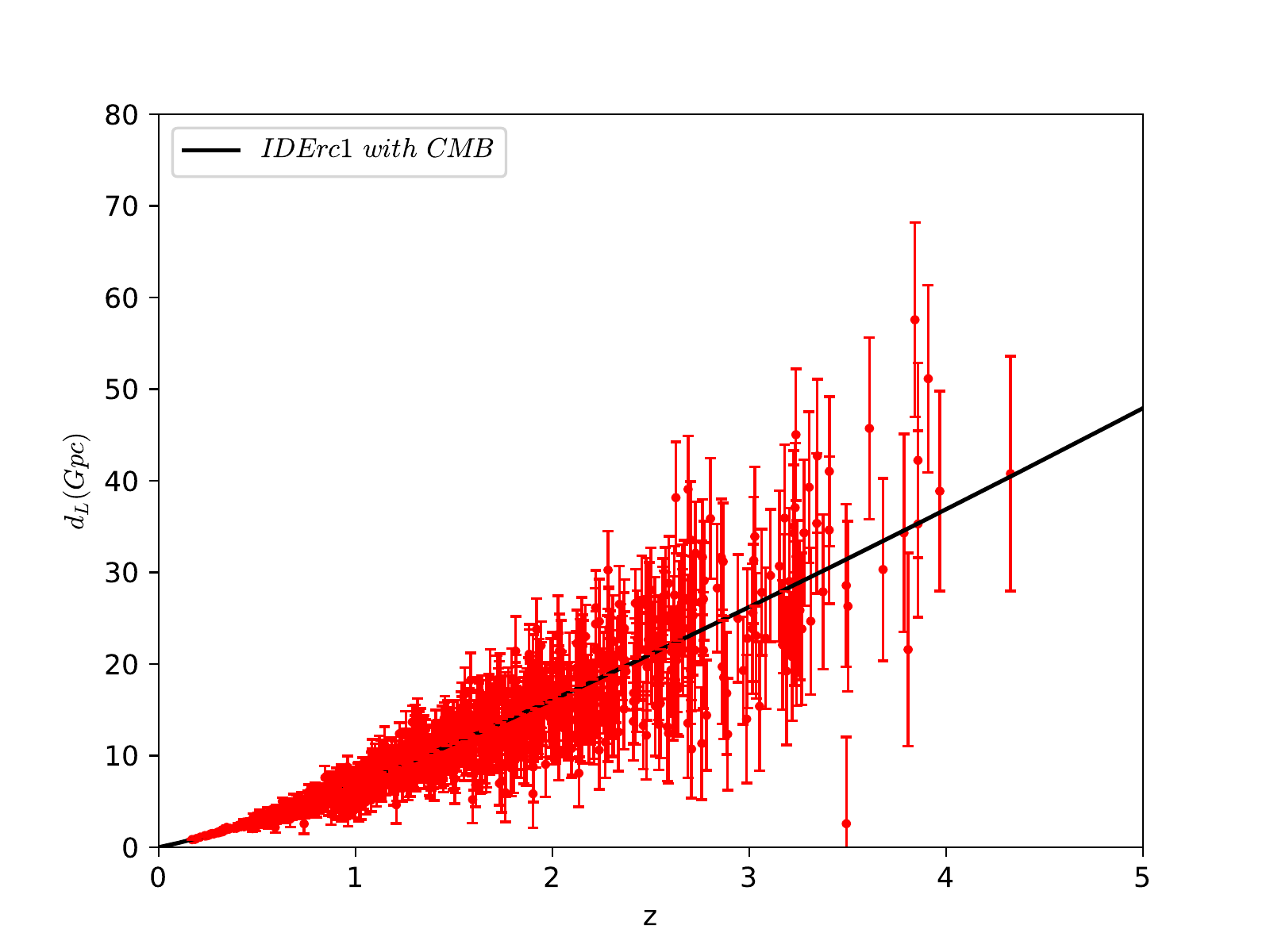}
\includegraphics[width=0.4\textwidth]{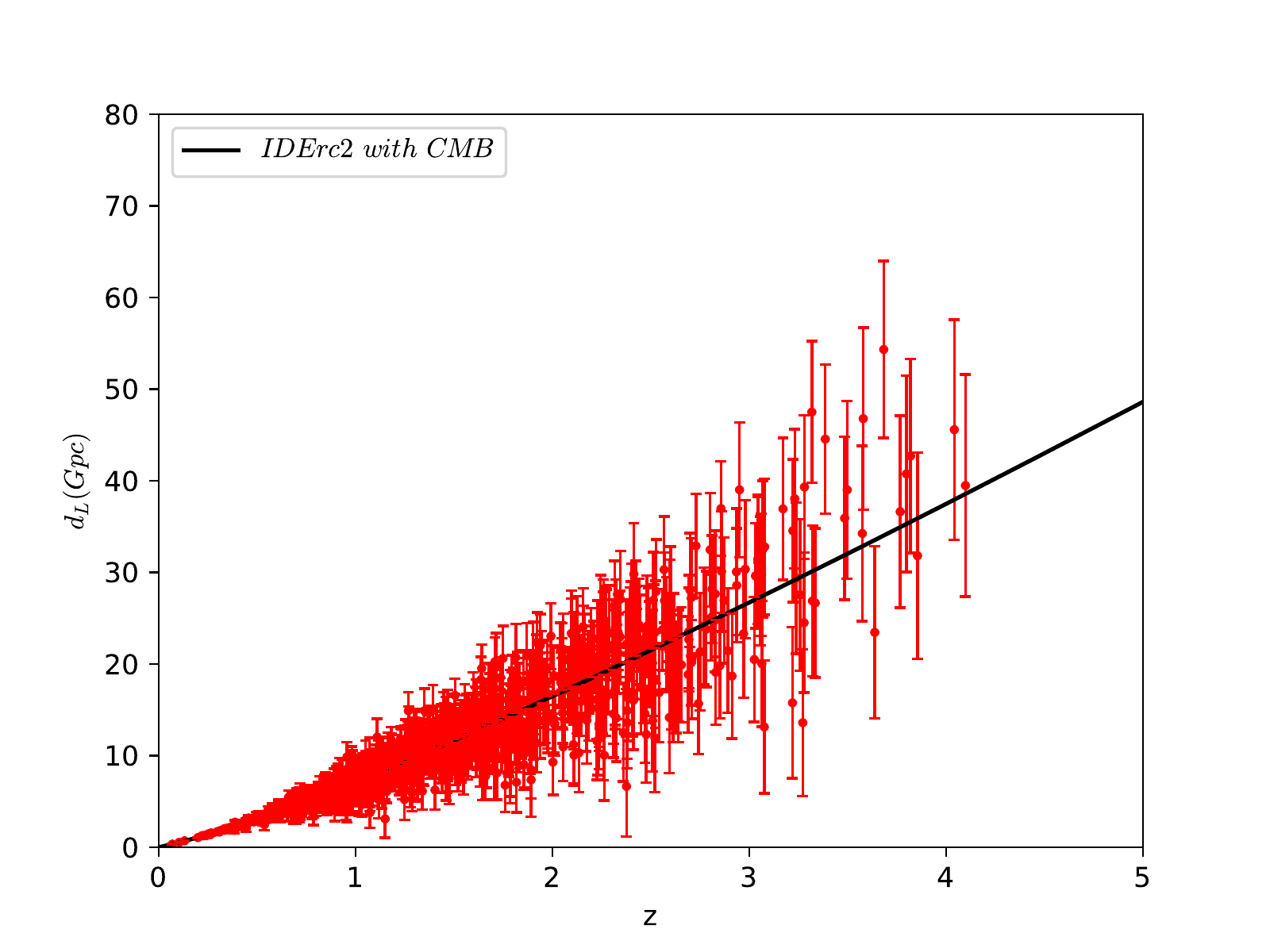}
\caption{Mock $d_L(z)$ measurements resulting from 1000 simulated GW events assuming a fiducial \texttt{IDErc1} model (left panel) and \texttt{IDErc2} model (right panel). The fiducial cosmological parameters are the best-fit values obtained when constraining these models against CMB data alone, except for $\xi$ which is fixed to $\xi=0.01$ and $\xi=0.025$ for the \texttt{IDErc1} and \texttt{IDErc2} models respectively.}
\label{fig:GW_luminosity}
\end{figure*}

This concludes the generation of the GW SS luminosity distance catalogue $(z_i,d_L(z_i),\sigma_i)$. In Fig.~\ref{fig:GW_luminosity} we show the mock GW data generated assuming as fiducial models the \texttt{IDErc1} (left) and \texttt{IDErc2} (right) models, with the black solid curves representing the theoretical prediction for $d_L(z)$. We model the GW likelihood as a product of Gaussians in $d_L(z_i)$ (with standard deviation $\sigma_i$), one Gaussian for each GW SS distance measurement in our mock catalogue. Therefore, we have modelled the 1000 events as being independent, and have neglected the cross-covariance between these measurements. More accurate analyses once real data is available should take this cross-covariance into account, as well as possible modelling systematics. For the time being, we have followed a more simple treatment and neglected these effects, which we will return to in future work.

We analyse two dataset combinations: CMB and CMB+GW. As previously stated, we generate the mock GW data using the best-fit values of the cosmological parameters from the CMB-only analysis. The only exception is the DM-DE coupling $\xi$ (whose best-fit value would nominally be $\xi=0$). We take the fiducial value of $\xi$ to be $\xi=0.010$ and $\xi=0.025$ when generating the GW data for the \texttt{IDErc1} and \texttt{IDErc2} models respectively. Both these values are within the $68\%$ confidence level (C.L.) upper limits obtained analysing the CMB dataset. The reason is that we want to check whether a non-zero coupling currently allowed by CMB data at $68\%$~C.L. is potentially discernible from zero coupling once GW SS data is added. In other words, we want to understand whether the addition of GW SS data can improve the uncertainty on $\xi$ to the point that we can detect non-zero $\xi$, should the true model chosen by Nature really feature dark sector interactions with a strength allowed by current data.

Finally, at a later stage we consider Baryon Acoustic Oscillations (BAO) distance measurements from the 6dF Galaxy Survey~\cite{Beutler:2011hx}, the Main Galaxy Sample of Data Release 7 of the Sloan Digital Sky Survey~\cite{Ross:2014qpa}, and the CMASS and LOWZ samples of Data Release 12 of the Baryon Oscillation Spectroscopic Survey~\cite{Gil-Marin:2015nqa}. We refer to this dataset as ``BAO''. Since BAO data is usually used in combination with CMB data to break parameter degeneracies, our goal is to check whether the CMB+GW combination will yield parameter constraints comparable to, or better than, the CMB+BAO combination. We further discuss our use of BAO data in Appendix~B, where we also provide a more detailed description of the BAO measurements.

In principle, we could also have chosen to include type-Ia Supernovae (SNeIa) luminosity distance measurements. One could in fact expect that the most important gain in combining GW and SNeIa distance measurements would be the fact that the two suffer from completely different systematic uncertainties. However, given that in our use of GW data we have not attempted to model systematic uncertainties, it is unlikely that such a gain would be appreciable in our analysis. For this reason, and in the interest of conciseness, we have chosen not to include SNeIa measurements.

We work within the framework of a 8-dimensional $\Lambda$CDM+$w_x$+$\xi$ cosmological model, described by the usual 6 $\Lambda$CDM parameters (the baryon and CDM physical densities $\Omega_bh^2$ and $\Omega_ch^2$, the angular size of the sound horizon at recombination $\theta_s$, the amplitude and tilt of the primordial power spectrum of scalar fluctuations $A_s$ and $n_s$, and the optical depth to reionization $\tau$), the DE EoS $w_x$, and the DM-DE coupling $\xi$. We impose flat priors on all these parameters unless otherwise stated, with prior ranges shown in Tab.~\ref{tab_priors}. Note that $\xi>0$ is required for the perturbations of both the \texttt{IDErc1} and \texttt{IDErc2} models to be stable~\cite{Yang:2018euj}. We sample the posterior distribution of the parameter space using Markov Chain Monte Carlo (MCMC) methods, and generating MCMC chains through the publicly available MCMC sampler \texttt{CosmoMC}~\cite{Lewis:2002ah}. The convergence of the generated chains is monitored through the Gelman-Rubin statistic $R-1$~\cite{Gelman-Rubin}.

\begin{table}[tbp]
\begin{center}
\begin{tabular}{c|c c}
Parameter & Prior range   \\ \hline
$\Omega_bh^2$ & $[0.005,0.1]$   \\
$\Omega_ch^2$ & $[0.01,0.3]$   \\
$\tau$ & $[0.01,0.8]$   \\
$n_s$ & $[0.5, 1.5]$  \\
$\log[10^{10}A_{s}]$ & $[2.4,4]$   \\
$100\theta_s$ & $[0.5,10]$   \\
$w_x$ & $[-2,0]$   \\
$\xi$ & $[0,2]$ 
\end{tabular}%
\end{center}
\caption{Prior ranges imposed on the 8 cosmological parameters of the interacting DE-DM models considered in this work.}
\label{tab_priors}
\end{table}
\begin{figure*}
\centering
\includegraphics[width=0.6\textwidth]{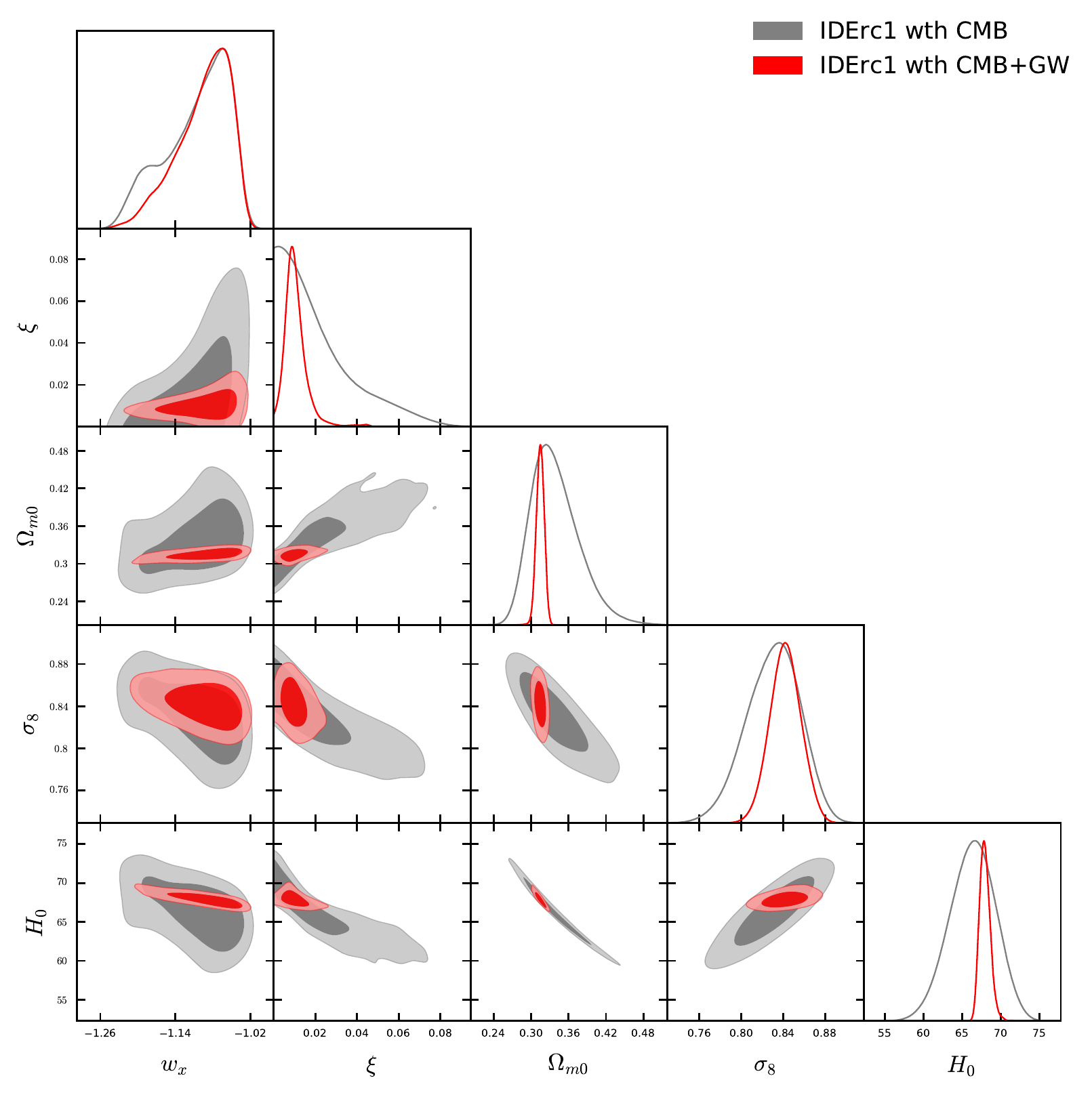}
\caption{1D marginalized and 2D joint posterior distributions for selected parameters (including a selection of derived parameters) of the \texttt{IDErc1} model whose determination is particularly improved by the inclusion of the GW dataset: $w_x$, $\xi$, $\Omega_{m0}$ (the total matter density parameter today), $\sigma_8$, and $H_0$ (in ${\rm km}\,{\rm s}^{-1}\,{\rm Mpc}^{-1}$). Contours are obtained using only CMB data (grey) and CMB+GW data (red).}
\label{fig:IDErc1}
\end{figure*}
\section{Results}
\label{sec:results}

We now examine the improvements in the constraints on the IDE parameters (especially the DM-DE coupling $\xi$) within the \texttt{IDErc1} and \texttt{IDErc2} models, brought upon by the addition of GW SS data. As described in Sec.~\ref{sec:data}, initially we constrain the parameters of the two IDE models using CMB data alone. Using the obtained best-fit values for all cosmological parameters (except the DM-DE coupling $\xi$, see above), we generate two mock GW SS catalogues (one for each IDE model). We finally combine CMB and GW data and examine the improvement on the uncertainties of the cosmological parameters with respect to the case where only CMB data is used, and with respect to the case where the CMB+BAO dataset combination is considered.

\subsection{Results for model \texttt{IDErc1}}
\label{results-IDErc1}

In Tab.~\ref{tab:IDErc1}, we report observational constraints on the parameters of the \texttt{IDErc1} model from the CMB and CMB+GW dataset combinations. In Fig.~\ref{fig:IDErc1} we instead display a triangular plot showing the 1D marginalized and 2D joint posterior distributions for selected parameters (including a selection of derived parameters) whose determination is particularly improved by the inclusion of the GW dataset: $w_x$, $\xi$, $\Omega_{m0}$ (the total matter density parameter today), $\sigma_8$, and $H_0$. Fig.~\ref{fig:IDErc1} also shows the correlations/degeneracies between these parameters and how some of these degeneracies are broken by the addition of GW data.

The introduction of GW data clearly leads to a substantial improvement in the determination of certain parameters, such as $\Omega_{m0}$ and $H_0$, whose error bars have been reduced by about a factor of $5$. In fact, while the CMB alone determines $H_0$ and $\Omega_{m0}$ with an accuracy of $5\%$ and $11\%$ respectively, we expect an improvement reducing these uncertainties down to $1\%$ on $H_0$ and $2\%$ on $\Omega_{m0}$ with the addition of GW SS data. This improvement is not surprising, as within a flat Universe and within the redshift range probed by GW SS data, the background expansion and hence the luminosity distance-redshift relation is mostly governed by $\Omega_{m0}$ and $H_0$. The same is true for $\xi$, as a small amount of energy transfer between DM and DE is sufficient to alter the background expansion by an appreciable amount. From Tab.~\ref{tab:IDErc1}, we see that it should be possible to detect a non-zero $\xi=0.01$ at a significance of about $2\sigma$. From Fig.~\ref{fig:IDErc1}, the improvement in the determination of $\xi$ after marginalizing over the other cosmological parameters is appreciable. This improved determination is particularly helped by the fact that the $\xi$-$\Omega_{m0}$ and $\xi$-$H_0$ degeneracies are almost broken by the addition of GW data. Perhaps surprisingly, there is only a slight improvement in the determination of the DE EoS $w_x$.

From Fig.~\ref{fig:IDErc1}, it is clear that the power of GW SS data goes beyond solely an improvement in the determination of background quantities. In fact, we see that the inclusion of GW SS data has halved the uncertainty on $\sigma_8$, which probes the growth of structure. At a first glance this might seem surprising, since $\sigma_8$ cannot be ``directly'' probed by background measurements. In fact, the improvement in the determination of $\sigma_8$ is mostly ``indirect'': a better determination of background quantities which are strongly degenerate with $\sigma_8$ will naturally lead to an improved determination of the latter. From Fig.~\ref{fig:IDErc1}, we see that the improvement in the determination of $\Omega_{m0}$ is particularly helpful in this sense, since the introduction of GW SS data has almost completely broken the $\sigma_8$-$\Omega_{m0}$ degeneracy.

\begin{figure*}
\centering
\includegraphics[width=0.6\textwidth]{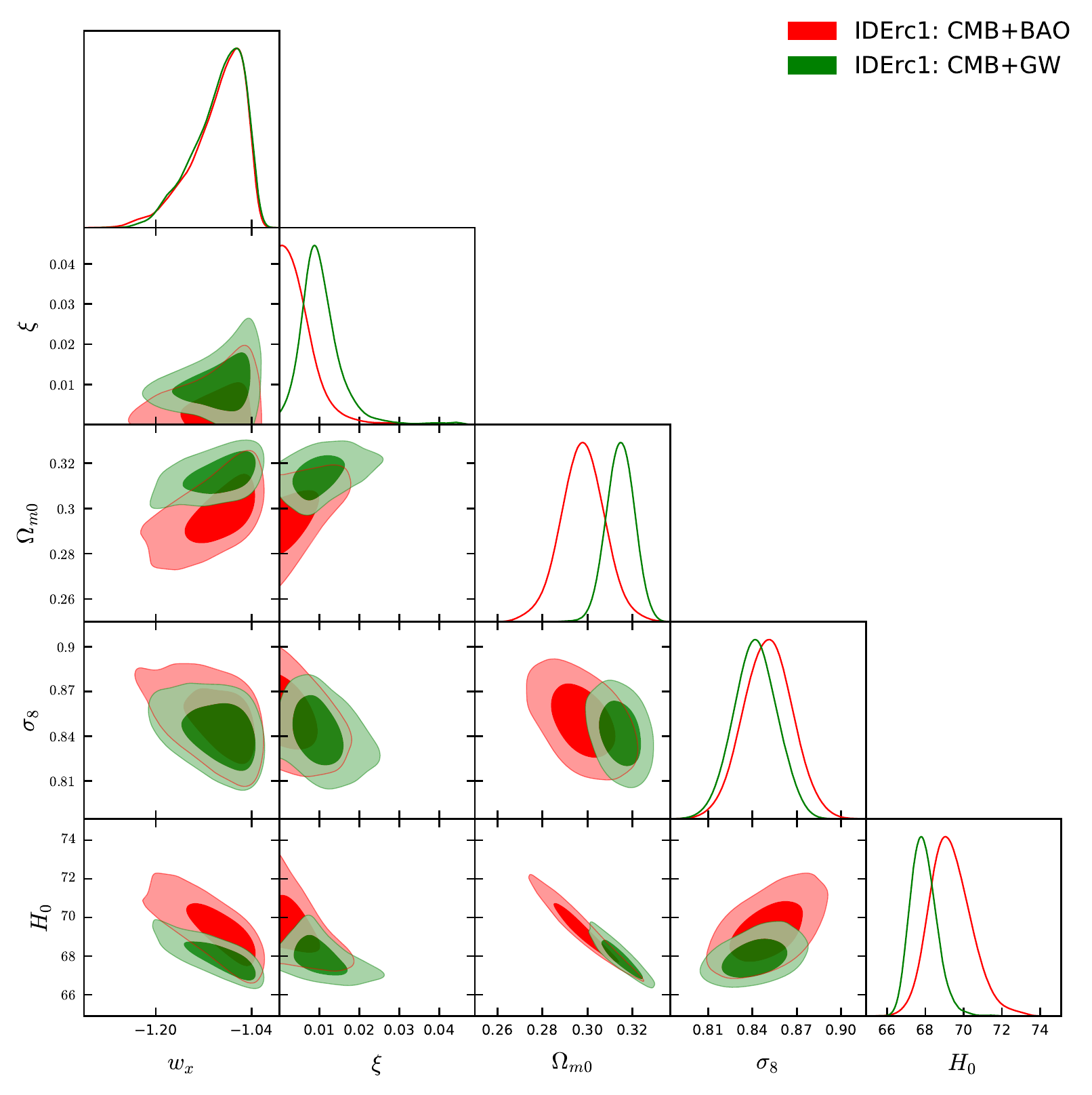}
\caption{1D marginalized and 2D joint posterior distributions comparing the CMB+BAO case (red contours) with CMB+GW (green contours), for selected parameters (including a selection of derived parameters), assuming the \texttt{IDErc1} model.}
\label{fig:IDErc1_BAO}
\end{figure*}

In Fig.~\ref{fig:IDErc1_BAO} we can compare the constraints obtained combining the CMB first with BAO measurements, and then with GW SS data. We can see that the CMB+GW combination will perform as well as the CMB+BAO case for most of the parameters, and will improve by about a factor of $2$ the determination of $H_0$ and $\Omega_{m0}$. For example, for the CMB+BAO case we have $\sigma_{H_0} \sim 1.1\,{\rm km}\,{\rm s}^{-1}\,{\rm Mpc}^{-1}$ and $\sigma_{\Omega_{m0}} \sim 0.01$, while we forecast $\sigma_{H_0} \sim 0.6\,{\rm km}\,{\rm s}^{-1}\,{\rm Mpc}^{-1}$ and $\sigma_{\Omega_{m0}} \sim 0.006$ for CMB+GW.
\begingroup         
\squeezetable       
\begin{center}       
\begin{table}
\begin{tabular}{ccccccccccccccc} 
\hline\hline                                        
Parameters & CMB &  CMB+GW \\ \hline
$\Omega_c h^2$ & $    0.125_{-    0.004-    0.007}^{+    0.003+    0.007}$ & $    0.122_{-    0.001-    0.002}^{+    0.001+    0.003}$ \\

$\Omega_b h^2$ & $    0.0223_{-    0.0002-    0.0003}^{+    0.0002+    0.0003}$ & $    0.0223_{-    0.0002-    0.0003}^{+    0.0002+    0.0003}$ \\

$w_x$ & $  -1.10_{-    0.03-    0.11}^{+    0.07+    0.09}$ & $   -1.10_{-    0.03-    0.09}^{+ 0.06+    0.07}$  \\

$\xi$ & $    0.022_{-    0.022-    0.022}^{+    0.004+    0.040}$ & $    0.011_{-    0.006-    0.010}^{+    0.003+    0.010}$  \\

$\Omega_{m0}$ & $    0.339_{-    0.047-    0.072}^{+    0.030+    0.081}$ & $    0.315_{-    0.006-    0.012}^{+    0.006+    0.012}$  \\

$\sigma_8$ & $    0.830_{-    0.025-    0.052}^{+    0.028+    0.051}$ & $    0.842_{-    0.015-    0.029}^{+    0.015+    0.029}$  \\

$H_0$ & $   66.2_{-    2.9-    5.9}^{+    3.3+    5.4}$ & $   67.9_{-    0.8-    1.4}^{+    0.6+    1.4}$  \\
\hline\hline
\end{tabular}
\caption{Observational constraints on selected cosmological parameters within the \texttt{IDErc1} model. Constraints on $H_0$ are reported in units of ${\rm km}\,{\rm s}^{-1}\,{\rm Mpc}^{-1}$ and $\Omega_{m0} = \Omega_b + \Omega_c$ is the total matter density today.}
\label{tab:IDErc1}                                
\end{table}
\end{center}
\endgroup          
However, it is worth commenting on a subtle point concerning the CMB+GW versus CMB+BAO comparison. This comparison is in reality not totally fair with regards to BAO measurements. It is in fact a comparison between measurements at completely different time scales: future GW data against current BAO data. The estimated ${\cal O}(10^3)$ GW SS events will only be available at the end of the Einstein Telescope science run, which will likely conclude between 2040 and 2060, a couple of decades from now. On the other hand, already within the next decade we will have sub-percent BAO measurements from a host of state-of-the-art large-scale structure surveys such as DESI~\cite{Levi:2013gra}, Euclid~\cite{Laureijs:2011gra}, and LSST~\cite{Abell:2009aa}. These measurements will considerably improve limits on cosmological parameters, including the DM-DE coupling $\xi$. One can safely extrapolate that such limits would surpass those of the CMB+GW combination in constraining power. Moving beyond these experiments, we do not yet know what the future of large-scale structure surveys has in store (especially not in 2060), but it is safe to say that the limits on dark sector interactions from BAO measurements on that time scale should considerably surpass those from the Einstein Telescope (unless the estimated ${\cal O}(10^3)$ GW SS events were somehow too pessimistic and we should serendipitously detect a significantly larger number of GW SS events). In any case, while we can certainly say that the addition of GW SS data to current CMB data will lead to important improvements in our constraints on dark sector interactions, one should keep in mind that the comparison with BAO data we have performed here is carried out between measurements living on completely different time scales.

\begin{figure*}
\centering
\includegraphics[width=0.6\textwidth]{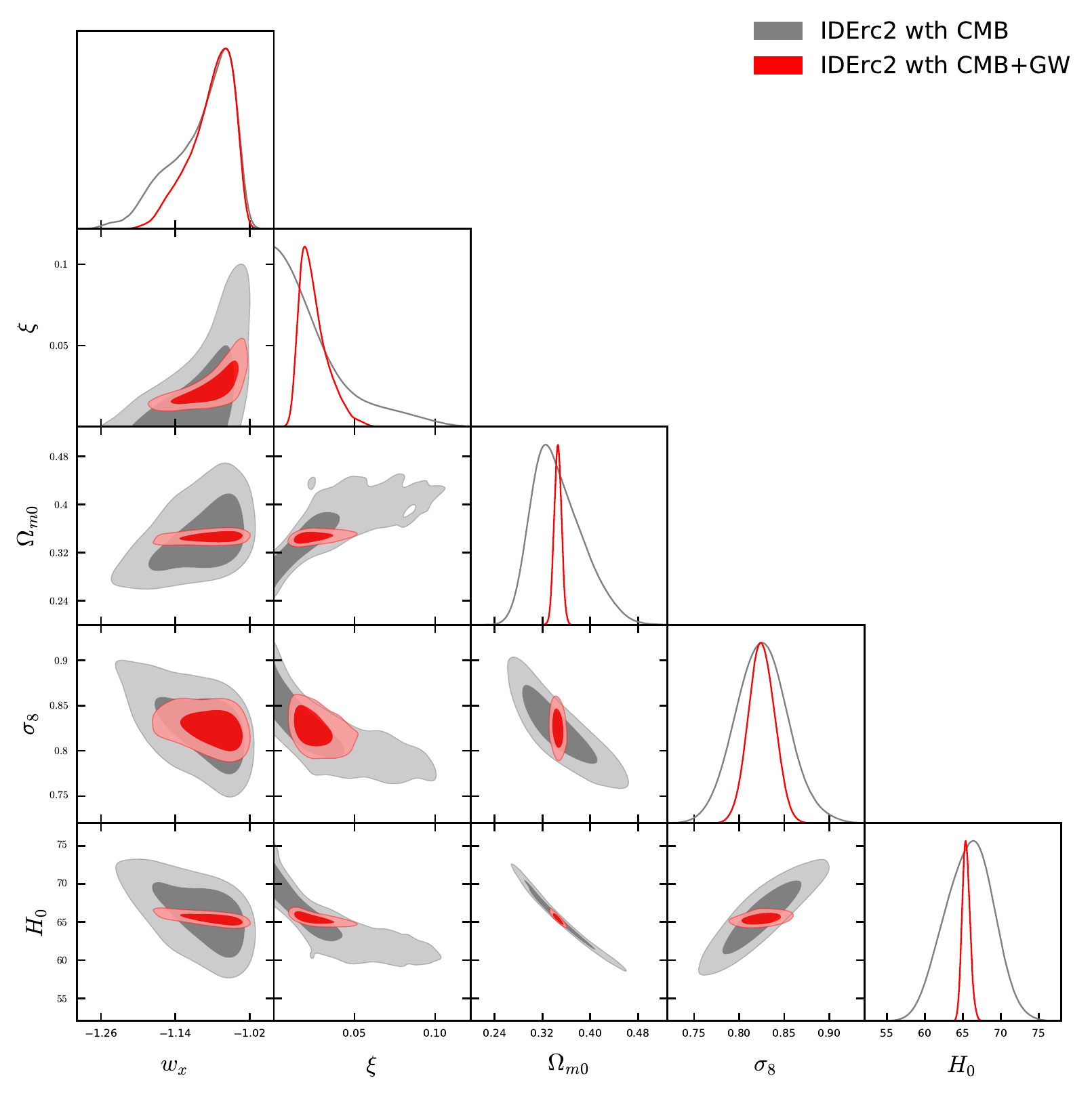}
\caption{As in Fig.~\ref{fig:IDErc1}, for the \texttt{IDErc2} model.}
\label{fig:IDErc2}
\end{figure*}

\subsection{Results for model \texttt{IDErc2}}
\label{results-IDErc2}

Our findings for the \texttt{IDErc2} model are summarized in Tab.~\ref{tab:IDErc2} and Fig.~\ref{fig:IDErc2}, which are completely analogous to their \texttt{IDErc1} counterparts. The results we find are extremely similar to those we found for the \texttt{IDErc1} model. In particular, we find that the introduction of GW data substantially reduces the error bars on quantities governing the background evolution such as $\Omega_{m0}$ and $H_0$, whose uncertainties are reduced by factors larger than $5$. In particular, while $H_0$ is determined with an accuracy of $5\%$ and $\Omega_{m0}$ of $12\%$ by the CMB, we predict an improvement up to $0.8\%$ on $H_0$ and $2\%$ on $\Omega_{m0}$ with the inclusion of the GW SS data.

Partially by breaking the $\xi$-$\Omega_{m0}$ and $\xi$-$H_0$ degeneracies, and also by better constraining the background evolution which is sensitive to $\xi$, GW data allow for a substantially more precise determination of $\xi$. In fact, we find that the $2\sigma$ uncertainty on $\xi$ has roughly halved after using GW SS data. In particular, we have found that it should be possible to detect $\xi=0.025$ at $>3\sigma$, \textit{i.e.} at relatively high significance. As for the \texttt{IDErc1} model, we find that there is only a marginal improvement in the determination of $w_x$, even if slightly larger in this case. On the other hand, we again find that the introduction of GW SS data halves the uncertainty on $\sigma_8$. Again, this is an ``indirect'' effect mostly brought upon by the fact that the $\sigma_8$-$\Omega_{m0}$ degeneracy is almost broken by the addition of GW data.

Finally, comparing the constraints obtained combining the CMB and BAO measurements versus CMB and GW SS data, we find results completely analogous to those obtained for the \texttt{IDErc1} model (which for conciseness we don't show here). That is, CMB+GW will perform better than CMB+BAO, improving also in this case considerably the determination of $H_0$ and $\Omega_{m0}$. However, we wish to remind the reader once more that the GW and BAO data we have considered live at completely different time scales (the final data from the Einstein Telescope is projected to be available between 2040 and 2060, whereas we have only considered current and not forecasted BAO data), and thus the comparison is perhaps not completely fair towards BAO measurements. Forecasts from next-generation BAO measurements (or even from BAO measurements available in 2060, although we currently have no idea what the projected sensitivity for BAO measurements will be at that point) would certainly lead to tighter constraints than those we have forecasted for the CMB+GW combination.
\begingroup         
\squeezetable       
\begin{center}      
\begin{table} 
\begin{tabular}{cccccccc} 
\hline\hline 
Parameters & CMB & CMB+GW \\ \hline
$\Omega_c h^2$ & $    0.126_{-    0.005-  0.007}^{+    0.003+    0.008}$ & $    0.125_{-    0.001-    0.002}^{+    0.001+    0.002}$  \\

$\Omega_b h^2$ & $    0.0223_{-    0.0002-    0.0003}^{+    0.0002+    0.0004}$ & $    0.0223_{-    0.0002-    0.0003}^{+    0.0002+    0.0003}$ \\

$w_x$ & $   -1.10_{-    0.03-    0.11}^{+    0.07+    0.08}$ & $   -1.08_{-    0.02-    0.08}^{+    0.05+    0.06}$  \\

$\xi$ & $    0.026_{-    0.026-    0.026}^{+    0.004+    0.050}$ & $    0.025_{-    0.011-    0.016}^{+    0.005+    0.020}$  \\

$\Omega_{m0}$ & $    0.346_{-    0.053-    0.077}^{+    0.033+    0.091}$ &  $    0.346_{-    0.006-    0.012}^{+    0.006+    0.012}$  \\

$\sigma_8$ & $    0.825_{-    0.029-    0.055}^{+    0.028+    0.058}$ & $    0.825_{-    0.014-    0.028}^{+    0.014+    0.028}$  \\

$H_0$ & $   65.8_{-    3.2-    6.1}^{+    3.5+    5.9}$ & $   65.4_{-    0.6-    1.0}^{+    0.5+    1.1}$  \\
\hline\hline      
\end{tabular}   
\caption{As in Tab.~\ref{tab:IDErc1}, for the \texttt{IDErc2} model.}
\label{tab:IDErc2}  
\end{table}          
\end{center}          
\endgroup 

\section{Summary and conclusions}
\label{sec:conclusions}

Interacting dark energy models, wherein dark matter and dark energy interact through couplings other than gravitational, have received renewed interest recently. While originally motivated by the possibility of therein addressing the coincidence problem, recently several of these models have been shown to potentially be able to address a number of discrepancies between high- and low-redshift determinations of cosmological parameters (\textit{e.g.} the $H_0$ tension and the $\sigma_8$ tension). Current observational datasets place relatively tight limits on the strength of the DM-DE coupling $\xi$ (with $95\%$~C.L. upper limits of order $\xi \lesssim 0.01$ depending on the IDE model under consideration), but are far from ruling out the possibility of DM-DE interactions.

With this in mind, our goal in this paper has been that of investigating how future distance measurements from GW standard sirens might improve constraints on IDE models. We considered two IDE models featuring stable early-time linear perturbations, originally proposed in~\cite{Yang:2018euj}. After generating ${\cal O}(10^3)$ mock GW standard sirens luminosity distance measurements matching the expected sensitivity of a 10-year run of the Einstein Telescope, a third-generation ground-based GW detector, we have studied how the addition of these GW distance measurements can improve current determinations of cosmological parameters within IDE models based on CMB data from \textit{Planck}. Our results show that the introduction of GW data is extremely helpful in pinning down background quantities such as $H_0$ and $\Omega_{m0}$. We find that GW standard sirens distance measurements can reduce the uncertainty on the DM-DE coupling $\xi$ by up to a factor of $5$ or more. We show that DM-DE interactions with strength within the current $1\sigma$ upper limit should be detectable at potentially more than $3\sigma$ with future GW data from the Einstein Telescope.

In conclusion, in this work we have demonstrated that future GW standard sirens distance measurements from the Einstein Telescope are expected to provide a powerful window onto the physics of dark sector interactions. Therefore, it might in principle just be a matter of time before we might be able to convincingly detect dark sector interactions, which the current discrepancies between high- and low-redshift cosmological probes might be hinting to. The methodology we have adopted in this paper is useful for analysing several cosmological models beyond interacting dark energy. It might be promising to adopt a similar methodology to investigate modified theories of gravity, although in such case one must take into account the fact that the waveform and hence the observed GW strain will be modified. It would also be interesting to investigate the sensitivity of future GW detectors other than the Einstein Telescope, in combination with future ground-based CMB surveys such as \textit{Simons Observatory}~\cite{Ade:2018sbj} and CMB-S4~\cite{Abazajian:2016yjj} as well as BAO measurements from future large-scale structure surveys such as DESI~\cite{Levi:2013gra}, Euclid~\cite{Laureijs:2011gra}, LSST~\cite{Abell:2009aa}. A more accurate modelling might potentially improve the impact of GW data \cite{Feeney:2018mkj} (or, conversely, a smaller number of GW events would be required to reach the sensitivity we have forecasted in this work). We plan to develop these and related issues in future work.

\section*{ACKNOWLEDGMENTS}
The authors wish to thank the referee for her/his very useful comments, which have helped us to improve the quality of the discussion. W.Y. acknowledges financial support from the National Natural Science Foundation of China under Grants No. 11705079 and No. 11647153. S.V. acknowledges support by the Vetenskapsr\r{a}det (Swedish Research Council) through contract No. 638-2013-8993 and the Oskar Klein Centre for Cosmoparticle Physics, and from the Isaac Newton Trust and the Kavli Foundation through a Newton-Kavli fellowship, and thanks the University of Michigan, where part of this work was conducted, for hospitality. E.D.V. acknowledges support from the European Research Council in the form of a Consolidator Grant with number 681431. S.P. acknowledges partial support from the Faculty Research and Professional Development Fund (FRPDF) Scheme of Presidency University, Kolkata, India. D.F.M. thanks the Research Council of Norway for their support. \\

\section*{Appendix A: further details on the generation of the mock GW standard sirens dataset}

In this Appendix, we provide further technical details on the well-known methodology used to generate the mock GW standard sirens dataset, which we briefly described in Sec.~\ref{sec:data}. We also encourage the reader to consult~\cite{Du:2018tia}. This method was used to obtain the uncertainties on the luminosity distance measurements associated to the binary black hole-neutron star and binary neutron star merger events.

During the coalescence phase, the GW waveform is very well described by the stationary phase approximation, wherein it takes the form:
\begin{eqnarray}
h(f) = Af^{-7/3}e^{i\Phi(f)}\,,
\label{eq:waveform}
\end{eqnarray}
where the amplitude $A$ is given by:
\begin{eqnarray}
A & = & \frac{1}{d_L} \sqrt{F^2_{+}(1+cos^2(\iota))^2 + 4F^2_{\times}cos^2(\iota))} \nonumber  \\
& \times & \sqrt{5 \pi/96} \pi^{-7/6} M^{5/6}_c\,.
\end{eqnarray}
In the above, $M_c$, $\iota$, $F_+$, and $F_x$ are respectively the redshifted chirp mass, angle of inclination of the binary orbital angular momentum with the line-of-sight, and two antenna pattern functions associated with the Einstein Telescope (see Eq.~(18) in~\cite{Zhao:2010sz} for the exact functional form of the antenna pattern functions). Finally, $d_L(z)$ is the luminosity distance to the redshift of the merger, and is our physical observable of interest here. In Eq.~(\ref{eq:waveform}), the function $\Phi(f)$ is the inspiral phase of the binary system, computed perturbatively within the so-called post-Newtonian formalism. Here, we follow the standard assumption of assuming a correction up to 3.5 PN order given by the TaylorF2 waveform (for more details see the expansion of the coefficients in~\cite{Buonanno:2009zt}).

Once the waveform is well defined, the other relevant quantity for generating the mock catalogue is the signal-to-noise ratio (SNR) associated with each simulated event. The SNR is given by:
\begin{eqnarray}
{\rm SNR}^2 \equiv 4 Re \int_{f_{\min}}^{f_{\max}} df\, \frac{\vert h(f)\vert ^2}{S_n}\,,
\label{eq:snr}
\end{eqnarray}
where $S_n(f)$ is the spectral noise density of the Einstein Telescope detector (see Eq.~(19) in~\cite{Zhao:2010sz}). The upper cutoff frequency $f_{\max}$ is determined by the last stable orbit (isco), which marks the end of the inspiral regime and the onset of the final merger. We assume $f_{\max} = 2f_{\rm isco}$ Hz, where $f_{\rm isco} = 1/6^{3/2} 2 \pi M_c$. The lower cutoff frequency $f_{\min}$ is instead determined by the sensitivity of the Einstein Telescope, so we set $f_{\min} = 1\,{\rm Hz}$.

We are now ready to estimate the instrumental error on $d_L(z)$, which is given by $\sigma^{\rm ins}_{d_L} \simeq 2 d_L/{\rm SNR}$. The factor of 2 has been introduced to take into account the effect of the inclination angle for which the GW amplitude is maximum. On the other hand, GWs are lensed in the same way photons are, resulting a weak lensing effect error which we model as $\sigma^{\rm lens}_{d_L}= 0.05zd_L(z)$~\cite{Zhao:2010sz}. In doing so, we have not considered possible errors induced from the peculiar velocity due to the clustering of galaxies. Since our simulated events are at relatively high $z$, this is a safe assumption, as significant corrections due to peculiar velocities of galaxies are significant only for $z \ll 1$. In fact, at high $z$ the dominant source of uncertainty is the one due to weak lensing. Finally, the total uncertainty $\sigma_{d_L}$ on the luminosity distance measurements associated to each event are obtained by combining the instrumental and weak lensing uncertainties in quadrature, as follows:

\begin{eqnarray}
&\sigma_{d_L}& = \sqrt{ \left ( \sigma^{\rm ins}_{d_L} \right )^2 + \left ( \sigma^{\rm lens}_{d_L} \right )^2} \nonumber \\
&=& \sqrt{ \left ( \frac{2d_L(z)}{\rm SNR} \right )^2 + (0.05 z d_L(z))^2}\,.
\label{eq:sigmatot}
\end{eqnarray}

To conclude, as discussed in Sec.~\ref{sec:data}, when generating the luminosity distance measurements themselves (before even generating their uncertainties), one has to model the distribution of GW events. The redshift distribution of the sources, taking into account evolution and stellar synthesis, is well described by:
\begin{eqnarray}
P(z) \propto \frac{4 \pi \chi^2(z) R(z)}{H(z)(1+z)},
\end{eqnarray}
where $\chi$ is the comoving distance and $R(z)$ describes the time evolution of the burst rate and is given by $R(z)=1 + 2z$ for $z<1$, $R(z) = 3/4(5-z)$ for $1 \leq z \leq 5$, and $R(z)=0$ for $z>5$.

Finally, the input masses of the neutron stars and black holes are randomly sampled from uniform distributions within $[1-2]~{\rm M}_{\odot}$ and $[3-10]~{\rm M}_{\odot}$ respectively. When generating our mock GW events, we have only considered mergers with ${\rm SNR}>8$.

\section*{Appendix B: A comment on BAO data}

In this Appendix, we provide further details on the BAO measurements we used throughout the work. We use both isotropic (for~\cite{Beutler:2011hx} and~\cite{Ross:2014qpa}) and anisotropic (for~\cite{Gil-Marin:2015nqa}) measurements of the BAO scale. In general, BAO data measure the ratio between a distance scale $D$ and the length of a standard ruler $r_s$. What exactly the distance scale is depends on whether the measurement is isotropic or anisotropic, and we will return to this point later. As for the standard ruler, we make the assumption that $r_s$ coincides with the sound horizon at baryon drag, usually denoted by $r_d(z_d)$. This assumption is strictly speaking only valid for $\Lambda$CDM~\cite{Shafer:2015kda,Verde:2016ccp,Tutusaus:2016orn,Tutusaus:2017ibk} and might be broken when considering more exotic models. However, this should not be an issue for the IDE models under consideration, given that the deviation from $\Lambda$CDM therein is small (especially given the size of the allowed values of $\xi$). Therefore, in our work we have made the assumption that the standard ruler for BAO measurements is given by the sound horizon at baryon drag.

We now comment more on the distance scale appearing in BAO measurements. An isotropic BAO measurement will constrain the ratio $D_V(z_{\rm eff})/r_s$ between the dilation scale $D_V$ (at the effective redshift of the survey $z_{\rm eff}$) and the length of the standard ruler $r_s$. The dilation scale as a function of redshift is given by:
\begin{eqnarray}
D_V(z) = \left [ (1+z)^2D_A^2(z)\frac{cz}{H(z)} \right ]\,,
\label{eq:dilation}
\end{eqnarray}
where $D_A(z)$ and $H(z)$ denote the angular diameter distance and Hubble rate at redshift $z$ respectively. On the other hand, anisotropic BAO measurements can separate correlations along and across the line-of-sight. In this case, correlations along the line-of-sight will constrain the product $H(z)r_s$, whereas correlations across the line-of-sight will constrain the ratio $D_A(z)/r_s$.

One final point worth mentioning is that BAO measurements are usually obtained by analysing the 2-point correlation function (or power spectrum, or both) of a large-scale structure tracer sample (for example luminous red galaxies). Computing the correlation function or the power spectrum requires a choice of fiducial cosmology, within which one can compute fiducial values for all the scales of interest: $D_V^{\rm fid}(z)$, $D_A^{\rm fid}(z)$, $H^{\rm fid}(z)$, and $r_s^{\rm fid}(z)$. Technically speaking, BAO measurements then actually constrain quantities such as $[D_V(z_{\rm eff})/r_s]/[D_V(z_{\rm eff})/r_s]^{\rm fid}$ (where of course the quantity $[D_V(z_{\rm eff})/r_s]^{\rm fid}$ is a constant once the fiducial cosmology is specified), and correspondingly for anisotropic measurements.

\end{document}